\documentstyle[11pt,gh2001-asp,twoside,epsf]{article}
\def\lesssim{{_ <\atop{^\sim}}}
\def\grtsim{{_ >\atop{^\sim}}}

\def\lesssim{{_ <\atop{^\sim}}}
\def\grtsim{{_ >\atop{^\sim}}}
\def\msun{\mbox{M$_\odot$}}
\def\fd{\mbox{f$_d$}}
\def\vdt{\mbox{(V$_d/$V$_t$)$_{2.2}$}}

\def\edcomment#1{\iffalse\marginpar{\raggedright\sl#1\/}\else\relax\fi}
\reversemarginpar

\begin{document}
\title{Formation and evolution of galaxy disks: what is wrong with CDM?}

\author{Vladimir Avila-Reese and Claudio Firmani}
\affil{Instituto de Astronom\'\i a-UNAM, A.P. 70-264, 04510 
M\'exico, D. F.}

\begin{abstract}
The density distribution of galaxy disks formed within 
$\Lambda$CDM halos is in rough agreement with observations. The 
luminous-to-dark matter ratio increases with the surface brightness 
(SB). The lowest SB models tend to be minimum disks, but the high SB 
models hardly attain the maximum disk solution. With the introduction 
of shallow cores in the halos, high SB models become maximum disks. 
The shallow cores also help to improve the inner density profiles 
of bulge-less low SB models and the zero-point of the Tully-Fisher 
relation. The models predict well 
this relation and its scatter, as well as the small correlation among 
the residuals of this and the luminosity-radius relation, in spite of 
the dependence of the rotation curve shapes on SB.  
 
\end{abstract}

\index{galaxies: formation --- galaxies: halos --- galaxies: kinematics
and dynamics ---  galaxies: spiral --- dark matter}

\section{Introduction}

According to the Cold Dark Matter (CDM) scenario, disk galaxies 
form within hierarchically growing dark halos. In the last years, 
high-resolution N-body simulations and semi-analytical approaches 
allowed to understand the structure, correlations, and evolution 
of the CDM halos in such a way that the properties of the disks 
formed within them can be now modeled in detail and compared with 
observations. Several of these properties agree with observations 
when detailed angular momentum conservation, smooth gas accretion 
(no mergers), and negligible disk-halo feedback are assumed (e.g., 
Firmani et al. 1997; Dalcanton et al. 1997; Mo et al. 1998; 
Avila-Reese et al. 1998; van den Bosch 2000,2001; Avila-Reese \& 
Firmani 2000 (AF00); Firmani \& Avila-Reese 2000 (FA00); Buchalter
et al. 2001). 
Nevertheless, in more detail, potential difficulties seem to 
arise regarding e.g., the central luminous-to-dark matter ratios, 
the disk surface brightness profiles, and the zero-point of the 
Tully-Fisher relation.   

Here we explore these issues by using semi-numerical models, where the
coupled dynamics, hydrodynamics, SF and feedback of the 
dark/baryonic matter system are self-consistently solved under several
simplifying assumptions (see e.g., AF00).  
The virtue of the semi-numerical approach is that enables to follow 
the overall evolution of the halo/disk/bulge system (as in the 
numerical simulations) and, at the same time, allows to predict 
correlations and statistical properties of the galaxy population
(as in the semi-analytical models).

\section{Density distribution of galaxy disks}

{\bf The model.} The disk is build up within the gravitational 
potential of a growing
CDM halo. We use the extended Press-Schechter approach to generate the
statistical mass aggregation histories (MAHs) of the CDM halos, and a 
generalized secondary infall model to calculate the virialization of 
the accreting mass shells (Avila-Reese et al. 1998). The mass shells
are assumed to have aligned rotation axis and to be in solid body rotation, 
with specific angular momentum given by $j_{sh}(t_v)=dJ(t_v)/dM(t_v)$, 
where $J=\lambda GM^{5/2}/\left| E\right| ^{1/2}$, $J$, $M$ and $E$ are
the total angular momentum, mass and energy of the halo at the shell
virialization time $t_v$. The spin parameter, $\lambda$, is assumed to 
be constant in time (FA00). As the result of the assembling of these 
mass shells, a present day halo ends with an angular momentum distribution 
close to the universal distribution found in the N-body simulations by 
Bullock et al. (2001). A fraction \fd\ of the mass of each shell is assumed 
to cool down and form a disk in a dynamical time. The radial mass 
distribution 
of the infalling gas is calculated by equating its specific angular 
momentum to that of its final circular orbit in centrifugal equilibrium 
(detailed angular momentum conservation is assumed). The gravitational 
interaction of disk and halo is calculated 
using the adiabatic invariant formalism. 
The local SF within the growing disks is triggered by the 
Toomre gas gravitational instability criterion and self-regulated by 
a vertical disk balance between the energy input due to SNe and
gas accretion and the turbulent energy dissipation in the ISM. 
{\it The SF history in our models depends on the gas surface 
density determined mainly by $\lambda$, and on the gas accretion 
rate determined by the cosmological MAH}. Finally, we consider the 
formation of a secular bulge using the Toomre criterion for a stellar disk.  

\vspace*{0.3cm}

\noindent {\bf Results.} We obtain nearly exponential stellar 
surface density profiles, $\Sigma_{*}(r)$, with scale radii R$_*$ 
growing with time (inside-out formation). The question whether real
disks grow with time as in the hierarchical scenario is matter
of discussion currently. At $z=0$, the R$_*$ of the 
modeled disk galaxy population for a $\Lambda$CDM cosmology are in 
agreement with observations in the luminosity-radius (L-R) and 
velocity-radius diagrams. The gaseous disks are also roughly 
exponential with R$_g \sim 3-4$R$_*$.
The predicted relation between gas fraction and surface brightness (SB) 
and its scatter are in good agreement with observations (FA00; AF00). 
The surface density of the disks is determined mainly by 
$\lambda$ (see e.g., Fig. 3b in FA00); the mass and the MAH play a minor 
role: the disks are slightly less dense for smaller masses and/or more 
extended MAHs.

\begin{figure}
\vspace{5.5cm}
\includegraphics{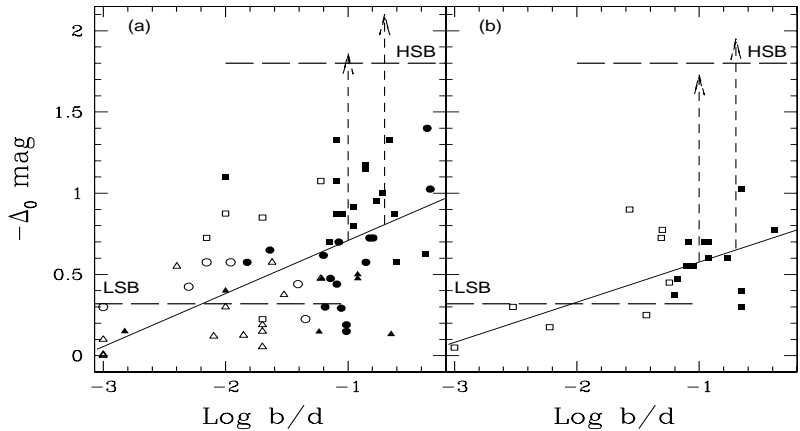}
\caption{{\it (a):} Central SB excess w.r.t. the exponential disk, 
$\Delta_0$, vs. b/d ratio of 60 random models in MAH and $\lambda$ 
and masses M$_v=3.5\times 10^{10}$ (triangles), 
$3.5\times 10^{11}$ (squares), and $3.5\times 10^{12} \msun$ (circles).
Arrows show by how much $\Delta_0$ would increase by the bar action
according to N-body simulations and the dashed lines are averages
for $\Delta_0$ from observations (see text). {\it (b):} The same
models of  M$_v=3.5\times 10^{11}\msun$ plotted in (a), but with shallow
cores in the halos.} 
\end{figure}

In more detail, the $z=0$ density profiles are typically more 
concentrated in the center and with an excess at the periphery than
the exponential distribution. Bullock et al. (2001) find that this is 
a direct consequence of the CDM halo $J-$distribution. 
We also find that $\Sigma_{*}(r)$ is mainly determined by the initial
$J-$distribution, but the sequential way in which the 
disk slabs aggregate (given by the halo MAH) and the way the gas is 
transformed in stars, also play a role. 
For the smoothed (averaged) MAHs with a given statistical
significance (see Fig. 1b in FA00), $\Sigma_{*}(r)$ always
presents a non-negligible core and tail excesses w.r.t. an exponential 
(Fig. 3a in FA00). However, when the individual realizations 
of the MAHs are used (Fig. 1a in FA00), in several cases the profiles 
result very close to an exponential law. The individual MAHs are 
discontinuous and some times ``jump'' from the high to the low 
accretion regimes and visceversa, so that using a simple 
parametrization for them (e.g., van den Bosch 2001), get to lose a 
significant fraction of possible MAHs. Note that $j_{sh}$ is a 
function of the time at which the mass shell is incorporated
into the halo, i.e. it depends on the MAH.

In Fig. 1 we show the central surface density (brightness) 
excess w.r.t. the exponential fit in magnitudes, 
$\Delta_{0}\equiv \mu_0 -\mu_{\rm 0,exp}$, vs. the bulge-to-disk mass 
(b/d) ratio that we obtain for disks formed within $\Lambda$CDM halos 
using randomly selected MAHs and $\lambda$ (the latter taken from the
usual lognormal distribution). The models are for a flat cosmology with 
$\Omega_{\Lambda}=h=0.7$ and a disk mass fraction \fd=0.05 (the same 
models as in FA00). 
The bulge mass is calculated as the mass where 
the Toomre instability criterion for a stellar disk is obeyed. This 
simple model is in line with the secular bulge formation scenario,
which seems to be adequate for disk galaxies later than Sba type 
(see Avila-Reese \& Firmani 1999 and the references therein). 
Gravitational instabilities in the disk produce bars which dissolve 
forming a bulge (Christodoulou et al. 
1995; Norman et al. 1996; Valenzuela \& Klypin 2002). Thus, according 
to the models, the region of core excess that disks formed within 
CDM halos present, is typically the region that will be transformed 
into a spherical component (AF00). 
  
In Fig. 1, estimates of the {\it average} $\Delta_{0}$ 
from the high SB (HSB) sample of disk galaxies of de Jong (1996) and 
Verheijen (1997) and from the low SB (LSB) sample of the latter author
are shown. 
These authors have decomposed the observed $K$-band SB profiles in 
(exponential) disk and bulge components. Solid and empty symbols in 
Fig. 1 are for models with disk central $\Sigma_{*,0}>200 \msun$ pc$^{-2}$ 
and $\Sigma_{*,0}\le 200 \msun$ pc$^{-2}$, respectively. Upon the 
understanding that the core excess $\Delta_{0}$ corresponds to a bulge, 
HSB models have in fact less concentrated SB profiles than observed 
HSB galaxies. Several processes could concentrate further the central 
disk/bulge region. For example, we did not take into account the mass 
inflow that will produce the bar before it dissolves into a bulge. 
High-resolution N-body simulations of a Milky Way disk within a CDM 
halo show that this process increases $\Sigma_{*,0}$ by $\approx 1.15$ 
and 1.30 mag for disks with initial R$_*$ of 3.5 and 3.0 kpc, 
respectively (Valenzuela \& Klypin 2002). These factors are plotted in 
Fig. 1 (dashed arrows) for the average $\Delta_0$ of the models 
corresponding to b/d=0.1 and 0.2, which are the final values
obtained by Valenzuela \& Klypin.

The LSB galaxy models have typically small or nonexistent bulges (models 
without bulges were plotted in Fig. 1 with b/d ratios of $10^{-3}$). In 
these cases, the SB core excess can not be interpreted at all as a bulge. 
Actually, there are LSB models with $\Delta_{0}$ below the observational 
average. Therefore, in some cases the pure disks formed within CDM halos
may be well fitted by an exponential law. This is at odds with the 
strong conclusion of van den Bosch (2001) that LSB disks formed
within CDM halos are much more concentrated than an exponential,
in conflict with observations. 
Although he implemented a disk/bulge formation scheme similar to ours,
he used only smoothed MAHs. When using these smoothed
MAHs, we obtain core excesses typically larger than when the 
random MAHs are used. We also differ with van
den Bosch (2001) in the way the SF is modeled: we use a physical 
self-consistent model of SF.  
The models predict a Schmidt law with index $\lesssim 2$, which
changes slighlty with radius.

Back to Fig. 1, one sees that the average $\Delta_{0}$ for the LSB 
models is larger than observations. As mentioned below, 
several pieces of evidence suggest shallow cores in dark halos.
In Fig. 1b we show the same $3.5\times 10^{11} \msun$ models as
in Fig. 1a, but with a shallow core in their halos. The 
disks formed within these halos are slightly less concentrated in 
the center than disks formed within cuspy halos.   
Regarding the tail excess of $\Sigma_{*}(r)$ w.r.t. an exponential, 
again, using the random MAHs, there are cases when this excess is 
even reverted. The observed SB profiles also show a diversity of cases 
with tail excess or defect. Although a more quantitative comparison 
of models and observations is desirable, {\it the density distributions 
of disks formed within CDM halos seem to be not in 
strong conflict with observations, in particular if the halos have 
a soft core.}

\section{Luminous-to-dark matter ratios}

The shapes of the model rotation curves (RCs) are smooth, showing 
a weak conspiracy between the dark and luminous matter 
distributions. The higher the disk surface 
density (the smaller the $\lambda$), the steeper the declining shape 
of the RC (see Fig. 5 in FA00 and Fig. 3b here); this is in agreement 
with some observational evidence (Casertano \& van Gorkom 1990; Verheijen 1997). 
The same trend is observed with the disk mass fraction \fd. 
Regarding the RC decomposition, the dark component dominates even in the
central regions for models with $\lambda\grtsim 0.035$ and $\fd\approx 0.05$.
For $\lambda\lesssim 0.045$, the models disks have large surface densities
typically (HSB galaxies). 

\begin{figure}
\vspace{5.2cm}
\includegraphics{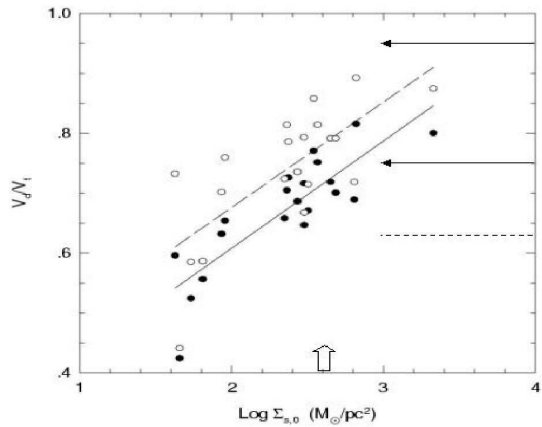}
\caption{Disk-to-total velocity ratio at 2.2 R$_*$ vs. $\Sigma_{*,0}$
of 20 random realizations in MAH and $\lambda$, and 
M$_v=3.5\times 10^{11} \msun$ for $\Lambda$CDM (filled circles) and 
shallow (empty circles) halos. Solid and long-dashed lines are linear 
regressions for the filled and empty circles. Arrows indicate the range 
of maximum disks, and the dashed line is for Bottema disks. A typical 
value of $\Sigma_{*,0}$ for HSB disks is shown with the open arrow.}
\end{figure}

The disk-to-total rotation velocity ratio at 2.2 scale radii, $\vdt\approx$
(V$_d$/V$_t$)$_{\rm max}$, 
offers a quantitative way to measure the model luminous-to-dark matter ratio.
In Fig. 2 we plot \vdt\ vs. $\Sigma_{*,0}$ for the 20 random realizations
with $M_v = 3.5\times 10^{11} \msun$ (solid symbols). The {\it 
luminous-to-dark matter ratio of disk galaxies within CDM halos 
continuously decreases from HSB to LSB disks}. Observations suggest
that HSB galaxies tend to the maximum disk case (e.g., Corsini et al. 
1998; Salucci \& Persic 1999; Palunas \& Williams 2000). Comparison of 
population synthesis models with the photometric properties of HSB 
galaxies also point to mass-to-luminosity ratios corresponding to 
the maximum disk case (e.g., Bell \& de Jong 2001). Theoretical arguments 
as the swing amplifier constraints (Athanassoula et al. 1987) also suggest 
central luminous matter dominion in HSB galaxies. Regarding LSB galaxies, 
it is well known that they are dark matter dominated systems (de Blok \& 
McGaugh 1997). A maximum disk solution for LSB galaxies demands too high 
mass-to-luminosity ratios from the stellar population point of view 
(de Blok et al. 2001).   

 Our results enable to interpret the problem of whether galaxies are 
maximum or sub-maximal disks: {\it all the cases are possible, from 
maximum disk for the highest SB galaxies to Bottema and other sub-maximal 
solutions for the lower SB galaxies}. However, for the $\Lambda$CDM, 
\fd=0.05 models we study here, the average value of \vdt\ is $\sim 0.7$ 
for a typical HSB galaxy ($\Sigma_{*,0}\approx 400\msun$ pc$^{-2}$), 
while for the maximum disk solution that observations seem to favor 
at least in some cases, $\vdt=0.85\pm 0.1$. Therefore, HSB galaxies 
formed within CDM halos are too much dark matter dominated in the 
center so as to be maximum disks. This is because 
the inner mass distribution of the CDM halos is very concentrated. 
Observations of dark matter dominated galaxies (dwarfs and LSB) 
show that the halo inner density profile is indeed shallower 
than CDM predictions (see e.g., Bosma, this volume). We artificially 
flattened the inner density profiles of the growing CDM halos in such a 
way that at $z=0$ (taking into account the gravitational drag of the 
disk) the halos have shallow cores similar to those inferred for observed 
LSB galaxies. Results for these models are shown with empty 
circles in Fig. 3. Now, a typical HSB galaxy can be maximum disk. 
The scatter in Fig. 3 is due to the MAH.
\vspace*{0.3cm}

\noindent {\bf The infrared Tully-Fisher relation (TFR) and its scatter.} 
According to the models, this relation is a direct
consequence of the mass-velocity relation of the $\Lambda$CDM halos, 
which in the range of galaxy masses has a slope of $\approx 3.2$ 
(FA00). The zero-point of the model TFR is fainter by $\sim 0.6$ mag 
w.r.t. the Giovanelli et al. (1997) and Tully \& Pierce (2000) 
$I$-band TFRs, but is in agreement with other observational determinations. 
In the N-body/hydrodynamical simulations, 
the zero-point is much fainter than the observed (Navarro, this volume). 
This is because in these simulations disks end much more concentrated 
than observed, producing highly peaked rotation curves, i.e. large 
V$_m$. Our models with a shallow core in the halos (see above) 
give a TFR only $0.1-0.2$ mag fainter than in the the two first 
observational works mentioned above.

The scatter in the model TFR is produced mainly by $(i)$ the scatter 
in the halo concentrations given by the stochastic nature of the 
MAHs, and $(ii)$ the scatter in the disk SBs given by the dispersion 
of $\lambda$. FA00 have found that the latter is reduced significantly 
due to the SF efficiency dependency on surface density. As is seen in 
Fig. 3a, 
for a given mass, V$_m$ increases with SB (see also above). This 
introduces a large scatter in the total disk mass-V$_m$ relation, 
even larger than the scatter due to the halo concentrations. 
However, the higher the disk surface density, the more efficient the 
gas transformation in stars, so that M$_*$ or the infrared luminosity 
of disks of the same mass will be larger for the higher SB ones: the models 
shift along the main TFR! (Fig. 4, solid arrows). Therefore, {\it 
the TFR of HSB and LSB galaxies is nearly the same}, in agreement 
with observations.
Neither does the disk mass fraction \fd\ introduce 
significant scatter in the TFR: for larger (smaller) \fd, the luminosity 
will be larger (smaller) but V$_m$ will also be larger (smaller) 
in such a way, that the model shifts along the main TFR (Fig 3, dotted 
arrows). The average scatter in the model TFR is 0.36 mag, mainly 
produced by the dispersion in the halo concentrations (MAHs).

\vspace*{0.3cm}
 
\noindent {\bf Correlation among the residuals of the TF and L-R 
relations.}
Because of the lack of correlation of the scatter in the TFR with SB, one 
also expects a lack of correlation among the residuals of the TF and 
L-R relations. Observations show indeed a very small correlation among 
these residuals (Courteau \& Rix 1999). The last authors interpreted 
this result as an evidence of {\it large dark halo dominion in disk 
galaxies}. If dark matter dominates strongly, then variations in the disk 
density will not change significantly V$_m$ for a given 
mass, therefore deviations from the main TFR in velocity 
will not be correlated with deviations from the main L-R relation. 
This interpretation is valid when one uses the total disk mass M$_d$ 
instead of M$_*$ or L. 
We find that the average slope of the correlation among the residuals of 
the M$_d$-V$_m$ and M$_d$-R relations for 
disks formed within the cuspy CDM halos is
$\delta$lgV$_m$/$\delta$lgR$\approx -0.35$. Therefore, according
to the interpretation of Courteau \& Rix, even CDM halos would then 
be in conflict with observations. If a shallow core is introduced, then 
the slope is even steeper, $\approx -0.4$ (this slope is $-0.5$ for maximum 
disks). However, in order to compare it with observations, we have to use 
M$_*$ or luminosity instead of M$_d$. In this case the correlation  
disappears! (Fig. 8 in FA00). The explanation is the same as 
why the scatter of the TFR does not correlate with SB in spite of the fact that 
the RC shape does: the dependence of SF efficiency on surface
density makes the difference (see above). In more detail, we predict that 
from the small to the large $\delta$lgR side (from HSB to LSB disks), 
$\delta$lgV$_m$/$\delta$lgR is first negative, then 
becomes 0 (for most of the normal galaxies) and then increases for the 
lowest SB galaxies. A preliminary analysis from observations using the 
Verheijen (1997) sample confirms this trend (Fig. 8b in FA00). 

\begin{figure}
\vspace{6.3 cm}
\includegraphics{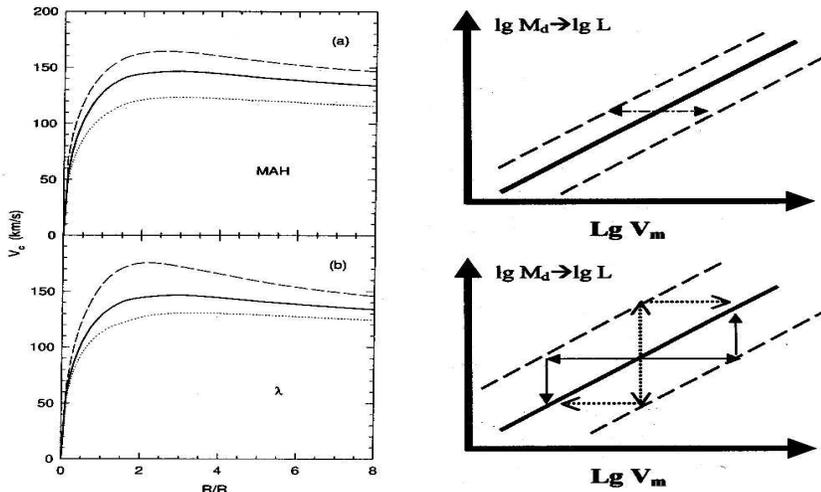}
\caption{Rotation curves for M$_v=3.5\times 10^{11} \msun$, \fd=.05,
models with 3 different MAHs and $\lambda=.05$ (a), and the average
MAH and $\lambda=.03, .05$ and .08 (b). The former introduces a scatter
in the halo concentrations that remains imprinted in the 
M$_d$-V$_m$ and TF relations (top). The latter introduces a 
scatter in the M$_d$-V$_m$ relation which correlates with SB 
(bottom), but when one passes from M$_d$ to L, the models shift along 
the main relation (solid arrows). Variations in \fd\ (dotted 
arrows), also shift models along the main relation (see text).}

\end{figure}

\section{Conclusions}

Using self-consistent evolutionary models of disk galaxy formation
within $\Lambda$CDM halos, and under the assumption of detailed angular 
momentum conservation and $\lambda=$const in time, we find that the 
{\it disk stellar and gas surface density profiles are in good 
agreement with observations.} The core excess w.r.t. the exponential 
law that the models show corresponds to a bulge formed by disk 
gravitational instabilities (AF00). In the case of bulge-less LSB disks, 
although there are models presenting negligible core excess, most 
of them tend to show slightly more concentrated SB profiles than those 
observed. The introduction of shallow cores in the halos alleviates 
this problem. A key point to have in mind is the very discontinuous 
nature of the cosmological MAHs and the SF physics.

A potential shortcoming
of disks formed within CDM halos is that dark matter dominates even for
HSB galaxies. Again, the introduction of shallow cores in agreement
with direct inferences from dwarf and LSB galaxies, solves this problem.
The models show that the luminous-to-dark matter ratio in disk galaxies
continuously decreases with SB: {\it the highest SB galaxies are maximum 
disks while the lowest SB ones are close to minimal disks}.  

The slope, zero-point and scatter of the model infrared TFR for HSB
and LSB galaxies are in good agreement with the observational reports 
of several authors. The introduction of shallow cores in the CDM 
halos improves the comparison. {\it Even that the RC shape of 
the models correlates with the disk SB, a negligible correlation among 
the residuals of the TF and L-V relations is found}, in agreement with 
observations. Therefore, the lack of this correlation should not be 
interpreted as a definitive evidence of sub-maximal disks, where the RC 
shape does not depend on SB. We conclude that $z=0$ disks formed within 
modified (shallow) CDM halos are realistic under the assumption of no 
major merger assembly and angular momentum conservation.

\acknowledgments V.A. thanks the organizers 
for financial help and for having achieved an interesting and fruitful 
meeting, and O. Valenzuela for helpful comments. This work was supported 
by CONACyT grant 33776-E to V.A.

\end{document}